\documentclass[a4paper]{article}

\usepackage{INTERSPEECH2019}
\usepackage{cite}
\usepackage{tablefootnote}
\usepackage{color}
\usepackage{pifont}
\usepackage{footnote}
\makesavenoteenv{tabular}

\usepackage{soul}

\def\X#1{%
        \raisebox{.9pt}{\textcircled{\raisebox{-.9pt}{#1}}}%
}
\newcommand{\quotes}[1]{``#1''}
\title{Improving Unsupervised Subword Modeling via Disentangled Speech Representation Learning and Transformation}
\name{Siyuan Feng, Tan Lee}
\address{
  Department of Electronic Engineering, The Chinese University of Hong Kong, Hong Kong}
\email{siyuanfeng@link.cuhk.edu.hk, tanlee@ee.cuhk.edu.hk}

\begin{document}
\maketitle
\begin{abstract}
This study tackles unsupervised subword modeling in the zero-resource scenario, learning frame-level speech representation that is phonetically discriminative and speaker-invariant, using only untranscribed speech for target languages. Frame label acquisition is an essential step in solving this problem. High quality frame labels  should be in good consistency with golden transcriptions and robust to speaker variation. We propose to improve frame label acquisition in our previously adopted deep neural network-bottleneck feature (DNN-BNF) architecture by applying the factorized hierarchical variational autoencoder (FHVAE). FHVAEs learn to disentangle linguistic content and speaker identity information encoded in speech. By discarding or unifying speaker information, speaker-invariant features are learned and fed as inputs to DPGMM frame clustering and DNN-BNF training. Experiments conducted on ZeroSpeech 2017 show that our proposed approaches achieve $2.4\%$ and $0.6\%$ absolute ABX error rate reductions in across- and within-speaker conditions, comparing to the baseline DNN-BNF system without applying FHVAEs. Our proposed approaches significantly outperform vocal tract length normalization in improving frame labeling and subword modeling.

\end{abstract}
\noindent\textbf{Index Terms}: unsupervised subword modeling, disentangled representation, speaker-invariant feature, zero resource

\section{Introduction}
Recent years have witnessed a huge success in applying deep learning  techniques in acoustic and language modeling for automatic speech recognition (ASR). 
Training  deep neural network (DNN) acoustic models  requires large amounts of transcribed speech data. 
For many languages in the world, for which very little or no transcribed speech is available, conventional supervised acoustic modeling techniques cannot be  directly applied.

Unsupervised acoustic modeling (UAM) aims at discovering and modeling acoustic units  of an unknown language at subword or word level, assuming only untranscribed speech data are available.
UAM is a challenging problem with significant practical impact in speech  as well as linguistics and cognitive science communities. It has been studied in applications  such as ASR for low-resource languages \cite{I3EWang}, language identification \cite{li2007vector} and query-by-example spoken term detection \cite{Chen+2016}. This problem is also relevant to
endangered language protection \cite{jansen2013summary} and  understanding infants' language acquisition mechanism \cite{dupoux2016cognitive}. 

Over the recent past, Zero Resource Speech Challenges (ZeroSpeech) 2015 \cite{versteegh2015zero} and 2017 \cite{dunbar2017zero} were organized to focus on unsupervised speech modeling.
ZeroSpeech 2017 Track one, named unsupervised subword modeling,  was formulated as an unsupervised feature representation learning problem, i.e., how to learn frame-level speech features that are discriminative to subword units and robust to linguistically-irrelevant variations such as speaker identity. 
The present study addresses this problem.
It is a fundamental problem in unsupervised speech modeling.
Speech simultaneously encodes linguistically-relevant information e.g. subword units and linguistically-irrelevant information e.g. speaker variation that are not easily separable. In supervised acoustic  modeling, golden transcription can be relied on to ensure the robustness of the learned subword units towards linguistically-irrelevant information. In the unsupervised scenario, subword units and word patterns can only be  inferred from speech features. This makes feature representation learning important in the zero-resource scenario. In the literature, representation learning has been shown beneficial to downstream applications such as spoken query retrieval \cite{chen2017multitask}.

In our previous attempt to ZeroSpeech 2017  \cite{Feng2018exploiting}, a  DNN was trained with   zero-resource speech data to generate bottleneck features (BNFs) as the learned feature representation. Frame labels for supervised DNN training were obtained through Dirichlet process Gaussian mixture model (DPGMM) based frame clustering. This framework is similar to \cite{chen2017multilingual}.
By employing  out-of-domain transcribed speech data for speaker adapted feature learning and DNN frame labeling, the results in \cite{Feng2018exploiting} significantly outperform  \cite{chen2017multilingual} in which out-of-domain data were not employed. 
This improvement is mainly attributed to the advancement of frame label acquisition.   
Ideally, the learned  frame labels should have a full coverage of linguistically-defined phonemes. They should be  in good consistency with golden transcription and  robust to speaker variation. The quality of frame labels has a significant impact on the performance of subword modeling \cite{heck2017feature}.
Many prior works found out that DPGMM clustering towards speaker adapted features could generate better labels than that towards unadapted features \cite{HeckSN16,heck2017feature,chen2017multilingual}. In \cite{chen2017multilingual}, the authors compared MFCC features with and without vocal tract length normalization (VTLN) for clustering. 
In \cite{heck2017feature}, MFCCs were first clustered to generate initial tokenization, with which  linear transforms such as LDA, MLLT and fMLLR  were estimated. The fMLLRs are clustered again to generate the final form of frame labels. This work achieved the best performance in ZeroSpeech 2017. 
It is worth noting that DPGMM clustering requires  high computational costs. 
Typically, clustering towards $40$-hour speech data for $100$ iterations using $32$ CPU cores takes up to $25$ hours. This makes the system in \cite{heck2017feature} much heavier than  \cite{chen2017multilingual,Feng2018exploiting}.

In the strict zero-resource scenario, out-of-domain speech and language resources are unavailable. 
This paper proposes to improve DPGMM  frame labeling using only in-domain untranscribed speech data, and refrain from performing multiple-pass clustering processes.
Specifically, the factorized hierarchical variational AE (FHVAE) model \cite{hsu2017nips} is used to disentangle linguistic content and speaker information in raw speech features in an unsupervised manner. By either discarding or unifying speaker information, speaker-invariant representation is learned and  used as the input to DPGMM clustering and DNN-BNF training. 
The FHVAE is an unsupervised generative model.
It was originally proposed to deal with domain adaptation problems in noise robust ASR \cite{hsu2018extracting}, distant conversational ASR \cite{hsu2018unsup},  and later applied to dialect identification \cite{shon2018unsup}. To the best of our knowledge, the use of FHVAEs in unsupervised subword modeling  has never been studied before. 

\section{Speaker-invariant feature learning by FHVAE}
Speaker characteristics tends to have a smaller amount of variation than linguistic content within a speech  utterance, while linguistic content tends to have similar amounts of variation within and across utterances. The FHVAE model \cite{hsu2017nips}, which learns to factorize sequence-level and segment-level   attributes of sequential data into different latent variables, 
is applied in this work to disentangle linguistic content and speaker characteristics.

\subsection{FHVAE model}
FHVAEs formulate the generation process of sequential data by imposing sequence-dependent priors and sequence-independent priors to different sets of variables. 
Following notations and terminologies in \cite{hsu2017nips}, 
let $\bm{z_1}$ and $\bm{z_2}$ denote latent segment variable and latent sequence variable, respectively. $\bm{\mu_2}$ is sequence-dependent prior, named as \emph{s-vector}. $\theta$ and $\phi$ denote the parameters of generation and inference models of FHVAEs.
Let $\mathcal{D}=\{\bm{X^{i}}\}_{i=1}^{M}$ denote a speech dataset with $M$ sequences. 
Each $\bm{X^i}$ contains $N^i$ speech segments $\{\bm{x^{(i,n)}}\}^{N^i}_{n=1}$, where $\bm{x^{(i,n)}}$ is composed of fixed-length consecutive 
frames. The FHVAE model generates a sequence $\bm{X}$ from a random process as follows: (1) $\bm{\mu_2 }$ is drawn from a prior distribution $p_{\theta}(\bm{\mu_2})=\mathcal{N} (\bm{0},\sigma^2_{\bm{\mu_2}} \bm{I})$; (2) $\bm{z_1 ^{n}} $ and $\bm{z_2^{n}} $ are drawn from $p_{\theta}(\bm{z_1 ^{n}})=\mathcal{N} (\bm{0}, {\sigma^2_{\bm{z_1}}} \bm{I})$ and  $p_{\theta}(\bm{z_2 ^{n}| \bm{\mu_2}})=\mathcal{N}(\bm{\mu_2}, {\sigma^2_{\bm{z_2}}} \bm{I} )$ respectively; (3) Speech segment $\bm{x^{n}}$ is drawn from $p_{\theta}(\bm{x^{n}}|\bm{z_1 ^{n}, \bm{z_2^{n}}})=\mathcal{N}(f_{\bm{\mu_x}} (\bm{z_1 ^{n}}, \bm{z_2^{n}}), diag(f_{\bm{\sigma^2_x}} (\bm{z_1 ^{n}}, \bm{z_2^{n}}))$. Here $\mathcal{N}$ denotes standard normal distribution, $ f_{\bm{\mu_x}} (\cdot, \cdot)$ and $ f_{\bm{\sigma^2_x}} (\cdot, \cdot)$ are parameterized by DNNs.
The joint probability for $\bm{X}$ is formulated as,
\begin{equation}
    p_{\theta} (\bm{\mu_2})\prod_{n=1}^{N} p_{\theta} (\bm{z_1^n}) p_{\theta} (\bm{z_2^{n}}|\bm{\mu_2})p_{\theta} (\bm{x^n}|\bm{z_1 ^{n}, \bm{z_2^{n}}}).
\end{equation}

Similar to VAE models, FHVAEs introduce an inference model $q_{\phi}$
to approximate the intractable  true posterior as,
\begin{equation}
   q_{\phi} (\bm{\mu_2})\prod_{n=1}^{N}q_{\phi} (\bm{z_2^n}| \bm{x^n}) q_{\phi}(\bm{z_1^n}|\bm{x^n}, \bm{z_2^n}).
   \label{eqt:inference}
\end{equation}
Here  $q_{\phi} (\bm{\mu_2}), q_{\phi} (\bm{z_2^n}| \bm{x^n})$ and $q_{\phi}(\bm{z_1^n}|\bm{x^n}, \bm{z_2^n})$ are all diagonal Gaussian distributions. The mean and variance values of $q_{\phi} (\bm{z_2^n}| \bm{x^n})$ and $q_{\phi}(\bm{z_1^n}|\bm{x^n}, \bm{z_2^n})$ are parameterized by two DNNs. For $q_{\phi} (\bm{\mu_2})$, during FHVAE training, a trainable lookup table containing posterior mean of $\bm{\mu_2}$ for each sequence is updated.
During testing, maximum a posteriori (MAP) estimation is used to infer  $\bm{\mu_2}$ for unseen test sequences. 
Details of $\bm{\mu_2}$ estimation for test sequences are described in \cite{hsu2017nips}.

FHVAEs optimize the discriminative segmental variational lower bound $\mathcal{L} (\theta, \phi; \bm{x^{(i,n)}})$ defined as,
\begin{align*}
&\mathbb{E}_{q_{\phi} (\bm{z_1^{(i,n)}},\bm{z_2^{(i,n)}}|\bm{x^{(i,n)}} )} [\log p_{\theta} (\bm{x^{(i,n)}}|\bm{z_1^{(i,n)}}, \bm{z_2^{(i,n)}})] -\\
& \mathbb{E}_{q_{\phi}(\bm{z_2^{(i,n)}} |\bm{x^{(i,n)}} )} [\mathrm{KL} (q_{\phi} (\bm{z_1^{(i,n)}}|\bm{x^{(i,n)}}, \bm{z_2^{(i,n)}})||p_{\theta} (\bm{z_1^{(i,n)}}))]\\
&-\mathrm{KL} (q_{\phi}(\bm{z_2 ^{(i,n)}}|\bm{x^{(i,n)}})|| p_{\theta}(\bm{z_2 ^{(i,n)}}| \bm{\tilde{\mu}_2^i})) \\
&+\frac{1}{N^i}\log p_{\theta} (\bm{\tilde{\mu}_2^i}) + \alpha \log p(i| \bm{z_2^{(i,n)}}),
\end{align*}
where $i$ is sequence index, $\bm{\tilde{\mu}_2^{i}}$ denotes posterior mean of $\bm{\mu_2}$ for the $i$-th sequence, $\alpha$ denotes the  discriminative weight. The discriminative objective $\log p(i| \bm{z_2^{(i,n)}}) $ is defined as $\log p_{\theta} (\bm{z_2^{(i,n)}}| \bm{\tilde{\mu}_2^i})-\log \sum_{j=1}^{M}p_{\theta} (\bm{z_2^{(j,n)}}| \bm{\tilde{\mu}_2^j})$.

After FHVAE training, $\bm{z_2}$ encodes factors that are relatively consistent within a sequence.
The discriminative objective ensures that $\bm{z_2}$ captures sequence-dependent information. $\bm{z_1}$ encodes residual factors that are sequence-independent.

\subsection{Extracting speaker-invariant features by FHVAE}
\label{subsec:spk_inv_feat}
In order to apply the FHVAE model to speaker-invariant feature learning, 
training utterances of the same speaker are concatenated into a single sequence.
By this means, 
$\bm{z_2}$ is expected to encode speaker identity information and  carry little phonetic information. $\bm{z_1}$ is expected to encode residual information, i.e. linguistic content, and carry little speaker information.
This work considers obtaining speaker-invariant feature representations based on a trained FHVAE by two methods. The first method is straightforward to treat 
latent segment variables $\{\bm{z_1^{(i,n)}}\}$ as the desired  feature representation. 

In the second method, the FHVAE model reconstructs speech features of all utterances based on a unified s-vector. The reconstructed features are the desired representation.
Specifically, a representative speaker with his/her s-vector $\bm{\mu_2^*}$ is chosen from the dataset. Next, for each speech segment $\bm{x^{(i,n)}}$ of an arbitrary speaker $i$, its corresponding latent sequence variable $\bm{z_2^{(i,n)}}$ is transformed to $\bm{\hat{z}_2^{(i,n)}}=\bm{z_2^{(i,n)}}-\bm{\mu_2^i}+\bm{\mu_2^*}$, where $\bm{\mu_2^{i}}$ denotes the s-vector of speaker $i$. 
Finally the FHVAE decoder reconstructs speech segment $\bm{\hat{x}^{(i,n)}}$ conditioned on $\bm{z_1^{(i,n)}}$ and $ \bm{\hat{z}_2^{(i,n)}}$ using $p_{\theta}(\bm{\hat{x}^{(i,n)}}|\bm{z_1 ^{(i,n)}, \bm{\hat{z}_2^{(i,n)}}})$. This method is named as \emph{s-vector unification} in this work.
Compared to original features,  reconstructed features  are expected to keep the  linguistic content unchanged and capture speaker characteristics corresponding to the representative speaker. In other words, speech synthesized from $\{\bm{\hat{x}^{(i,n)}}\}$ would tend to sound as if they were all spoken by the representative speaker. 

\section{Unsupervised subword modeling with speaker-invariant features}

\subsection{DNN-BNF architecture}
A DNN-BNF architecture  \cite{chen2017multilingual,Feng2018exploiting} is adopted to perform  phonetic discriminative training of untranscribed speech data and generate BNFs for subword modeling. 
In this  architecture, given untranscribed speech data,
Dirichlet process Gaussian mixture model (DPGMM) \cite{chang2013parallel} algorithm is applied to cluster frame-level MFCC features for each target language individually. 
After clustering, each frame is assigned with a cluster label. These frame labels  are regarded as pseudo phoneme alignments to support supervised DNN training. A multilingual DNN with a linear bottleneck layer is trained with frame alignments and MFCC features for all the target languages simultaneously, using multi-task learning  \cite{caruana1998multitask}. After training, multilingual BNFs are extracted as the subword discriminative  representation.

\subsection{DNN-BNF training with speaker-invariant features}
\begin{figure}[t]
    \centering
    \includegraphics[width=0.87\linewidth]{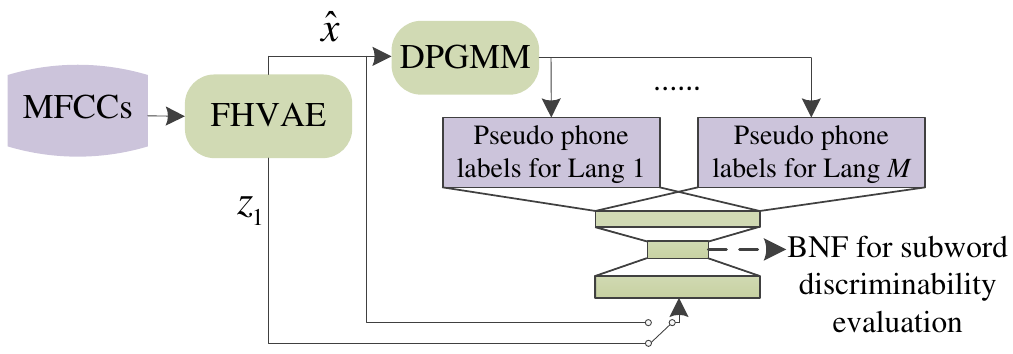}
    \caption{DNN-BNF architecture with FHVAE-based speaker-invariant features for unsupervised subword  modeling}
    \label{fig:framework}
\end{figure}

Speaker-invariant features learned by FHVAEs are applied  to the DNN-BNF architecture 
in two aspects.
As can be seen in Figure \ref{fig:framework}, during DPGMM-based frame clustering,  input features to DPGMM are reconstructed MFCCs $\{\bm{\hat{x}}\}$ generated by the FHVAE decoder network using the s-vector unification method described in Section 
\ref{subsec:spk_inv_feat}, instead of original MFCCs. Compared to original MFCCs, FHVAE reconstructed MFCCs  carry speaker information that is more consistent across utterances spoken by different speakers. With the reconstructed features as inputs, DPGMM clustering is expected to  generate better phoneme-like labels and less affected by speaker variation. 

During  DNN-BNF model training,  FHVAE-based speaker-invariant features are  fed as  inputs to the DNN. As seen in Figure \ref{fig:framework}, in this study we consider two feature types, i.e.  reconstructed MFCCs with s-vector unification $\{\bm{\hat{x}}\}$  and latent segment variables $\{\bm{z_1}\}$, as DNN inputs. The effectiveness of these two types of features  is compared in this study.  
\section{Experimental setup}
\subsection{Dataset and evaluation metric}
Experiments are carried out with ZeroSpeech 2017 Track one \cite{dunbar2017zero}.
Speaker identity information is released only for train sets. 
Detailed information  is listed in Table \ref{tab:zr17_data}.

\begin{table}[htbp]
\renewcommand\arraystretch{0.60}
\centering
\caption{Development data in ZeroSpeech 2017 Track one}
\resizebox{0.78 \linewidth}{!}{%
\begin{tabular}{lccc|c}      
\toprule
 & \multicolumn{3}{c|}{ Training} & Test \\
 & Duration  &\#speakers-R\tablefootnote{``speakers-R/-L'' denotes speakers with rich/limited speech data.} &\#speakers-L\footnotemark[1]  & Duration\\
\midrule
English & $45$ hrs &$9$& $60$ & $27$ hrs\\
French & $24$ hrs &$10$& $18$ & $18$ hrs\\
Mandarin & $2.5$ hrs &$4$& $8$ & $25$ hrs\\
\bottomrule
\end{tabular}%

}
\label{tab:zr17_data}
\end{table}

The evaluation metric is ABX subword discriminability. The ABX task is to decide whether $X$ belongs to $x$ or $y$ if $A$ belongs to $x$ and $B$ belongs to $y$, where $A$, $B$ and $X$ are three speech segments, $x$ and $y$ are two phonemes that differ in the central sound (e.g., \quotes{beg}-\quotes{bag}). 
Each pair of $A$ and $B$ are generated by the same speaker. 
ABX error rates for \textit{within-speaker} and \textit{across-speaker} are evaluated separately, depending on whether $X$ and $A(B)$ belong to the same speaker. 
\subsection{FHVAE setup and parameter tuning}
FHVAE model parameters are determined by reference to  \cite{hsu2018extracting}. The  encoder and decoder networks of FHVAE are both $2$-layer LSTMs with $256$ neurons per layer. The dimensions of  $\bm{z_1}$ and $\bm{z_2}$ are $32$. 
Training data for the three target languages are merged to train the FHVAE. 
Input features  are fixed-length speech segments randomly chosen from utterances. The determination of segment length $l$ is discussed in the next paragraph. Each frame is represented by a $13$-dimensional MFCC with cepstral mean normalization at speaker level. 
During the inference of reconstructed feature representation, input segments are shifted by $1$ frame. To match the length of extracted features with original MFCCs, the first and last frame are padded. 
Adam \cite{kingma2014adam} with $\beta_1=0.95$ and $\beta_2=0.999$  is used to train the FHVAE. A $10\%$ subset of training data is randomly selected for cross-validation.
The training process is terminated if the lower bound on the cross-validation set does not improve for $20$ epochs. Open-source tools \cite{hsu2017nips} are used to train FHVAEs.

\begin{figure}[t]
    \centering
    \includegraphics[width=0.92\linewidth]{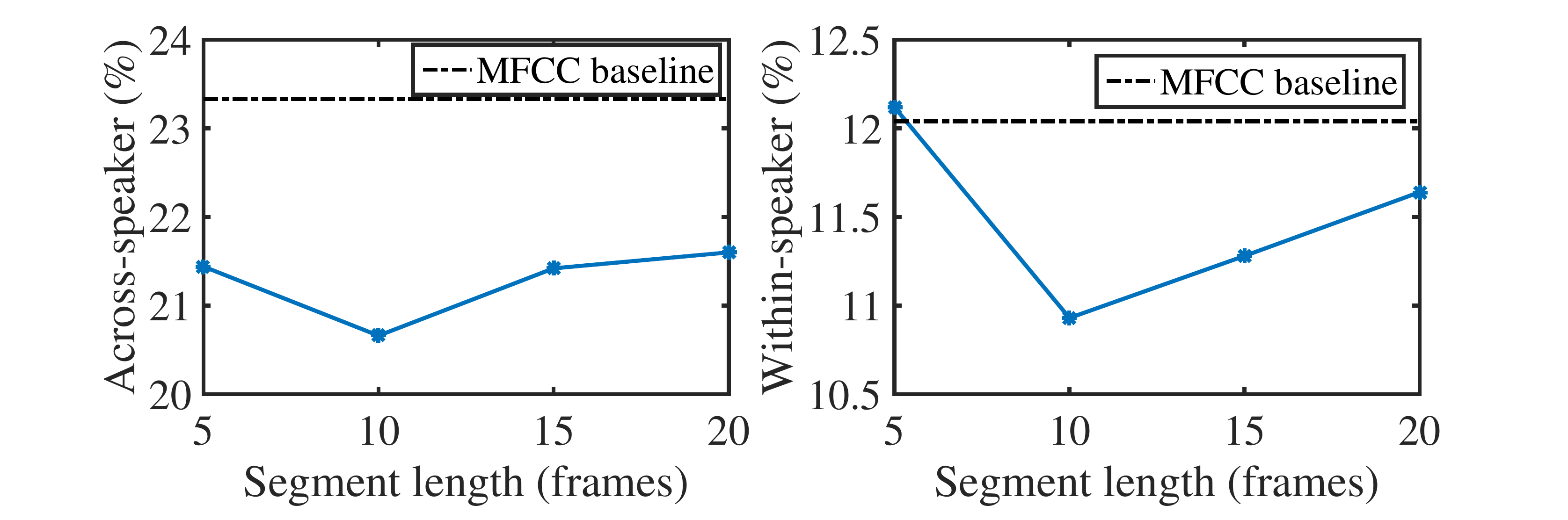}
    \caption{ABX error rates (\%) on  $\bm{z_1}$ with  different segment lengths and official MFCC baseline \cite{dunbar2017zero} (Avg. over languages)}
    \label{fig:tune_len}
\end{figure}

In our preliminary experiments, the ABX performance of $\bm{z_1}$ was found to be sensitive to the input segment length $l$. 
This could be explained as:  a too large  $l$ would reduce  the capability of $\bm{z_1}$ in modeling linguistic content at  subword level; 
a too small $l$ would restrict the FHVAE from capturing sufficient temporal dependencies which are essential in modeling speech.  ABX error rates on $\bm{z_1}$ with different values of $l$   are shown in Figure \ref{fig:tune_len}. 
The optimal value of $l$ is $10$. For the remaining experiments in this work, $l$ is fixed to $10$. 

\begin{table*}[htbp]
\renewcommand\arraystretch{0.54}
\centering
\caption{ABX error rates ($\%$) on DNN-BNFs trained with/without FHVAE-based speaker-invariant features}
\resizebox{ 0.92\linewidth}{!}{%
\begin{tabular}{cll|ccc|ccc|ccc|c||ccc|ccc|ccc|c}      
 \toprule[1pt]\midrule[0.3pt]
ID& && \multicolumn{10}{c||}{ Across-speaker} & \multicolumn{10}{c}{ Within-speaker} \\
\midrule
 & && \multicolumn{3}{c|}{ English} & \multicolumn{3}{c|}{ French} & \multicolumn{3}{c|}{Mandarin}& Avg.&\multicolumn{3}{c|}{ English} & \multicolumn{3}{c|}{ French} & \multicolumn{3}{c|}{Mandarin} & Avg.\\
&& & 1s & 10s & 120s & 1s & 10s & 120s & 1s & 10s & 120s && 1s & 10s & 120s & 1s & 10s & 120s & 1s & 10s & 120s\\ 
\midrule
&\multicolumn{2}{l|}{Baseline} &$13.5$&$12.4$&$12.4$&$17.8$&$16.4$&$16.1$&$12.6$&$11.9$&$12.0$&$13.90$&$8.0$&$7.3$&$7.3$&$10.3$&$9.4$&$9.3$&$10.1$&$8.8$&$8.9$&$8.82$ \\
 & \multicolumn{2}{l|}{CA-Sup \cite{Feng2018exploiting} }&$10.9$&$9.5$&$8.9$&$15.2$&$13.0$&$12.0$&$10.5$&$8.9$&$8.2$&$10.79$&$7.4$&$6.9$&$6.3$&$9.6$&$9.0$&$8.1$&$9.8$&$8.8$&$8.1$&$8.22$ \\
\midrule
&\multicolumn{2}{l|}{MFCC \cite{chen2017multilingual}} &$13.7$&$12.1$&$12.0$&$17.6$&$15.6$&$14.8$&$12.3$&$10.8$&$10.7$& $13.29$&
 $8.5$&$7.3$&$7.2$&$11.1$&$9.5$&$9.4$&$10.5$&$8.5$&$8.4$&$8.93$
 \\
&\multicolumn{2}{l|}{MFCC+VTLN \cite{chen2017multilingual}}& $12.7$&$11.0$&$10.8$&$17.0$&$14.5$&$14.1$&$11.9$&$10.3$&$10.1$&$12.49$&$8.5$&$7.3$&$7.2$&$11.2$&$9.4$&$9.4$&$10.5$&$8.7$&$8.5$&$8.97$
 \\
 \midrule
\X1 &$\bm{z_1}$& Orig. &  $12.9$&$11.7$&$11.7$&$17.2$&$15.5$&$15.2$&$12.5$&$11.4$&$11.5$&$13.29$&$8.2$&$7.0$&$7.0$&$10.7$&$9.2$&$9.1$&$10.4$&$8.8$&$8.7$&$8.79$ \\

\X2 &$\bm{\tilde{x}}$& Orig.& $12.8$&$11.7$&$11.5$&$17.8$&$15.5$&$15.1$&$12.3$&$10.9$&$10.7$&$13.14$&$8.2$&$7.3$&$7.0$&$10.6$&$9.3$&$8.9$&$10.5$&$8.8$&$8.7$&$8.81$ \\

\X3& $\bm{z_1}$&$\bm{\hat{x}}$-s0107& $11.2$&$10.1$&$10.1$&$15.5$&$13.8$&$13.7$&$11.5$&$10.2$&$10.0$&$11.79$&$7.3$&$6.4$&$6.6$&$10.1$&$8.9$&$8.8$&$10.4$&$8.5$&$8.4$&$8.38$ \\

\X4& $\bm{\tilde{x}}$&$\bm{\hat{x}}$-s0107& $11.6$&$10.4$&$10.1$&$16.1$&$13.9$&$13.7$&$11.9$&$10.2$&$10.4$&$12.03$&$7.8$&$6.7$&$6.5$&$10.5$&$9.6$&$9.3$&$10.8$&$8.6$&$8.7$&$8.72$ \\

\X5& $\bm{z_1}$&$\bm{\hat{x}}$-s4018 & $11.0$&$9.8$&$9.8$&$14.9$&$13.4$&$13.0$&$11.4$&$10.1$&$10.0$&$\bm{11.49}$&$7.3$&$6.3$&$6.3$&$9.7$&$8.6$&$8.4$&$10.1$&$8.5$&$8.4$&$\bm{8.18}$ \\

\X6& $\bm{\tilde{x}}$&$\bm{\hat{x}}$-s4018& $11.3$&$10.0$&$9.8$&$15.7$&$13.6$&$13.3$&$11.8$&$10.0$&$10.4$&$11.77$&$7.8$&$6.5$&$6.5$&$10.1$&$9.1$&$8.8$&$10.6$&$8.7$&$8.7$&$8.53$ \\
\midrule[0.3pt]\bottomrule[1pt]
\end{tabular}%
}
\label{tab:results}
\end{table*}
\subsection{Selecting representative speaker for reconstructed  feature extraction}
The extraction of reconstructed MFCCs $\{\bm{\hat{x}}\}$ using s-vector unification assumes a pre-defined representative speaker. 
In order to validate the generalization ability of our proposed s-vector unification method and evaluate its sensitivity to the gender of the representative speaker, $6$ English speakers \{s0107, s3020, s4018, s0019, s1724, s2544\}, $4$ French speakers \{M02R, M03R, F01R, F02R\} and $2$ Mandarin speakers \{A08, C04\} are randomly chosen from `speaker-R' sets of ZeroSpeech 2017 training data. 
The first half speakers  inside each language set  are male and the second half are female.
During the extraction of $\{\bm{\hat{x}}\}$, s-vectors  $\{\bm{\mu_2^i}\}$ of all three target languages' utterances are modified to the same $\bm{\mu_2^*}$  corresponding to one of the $12$ speakers mentioned above. The performance of  the $12$ groups of $\{\bm{\hat{x}}\}$ is evaluated by the ABX discriminability task.

\label{subsec:repre_spk}

\subsection{DNN-BNF setup}
For the baseline system without using FHVAE-based speaker-invariant features,  input features  to DPGMM are  $39$-dimensional MFCCs+$\Delta$+$\Delta\Delta$. 
The numbers of clustering iterations for English, French and Mandarin sets are $120,200$ and $3000$. After clustering, each frame is assigned with a label. A DNN-BNF is trained with all three languages'  cepstral mean normalized MFCCs+$\Delta$+$\Delta\Delta$ and frame labels using  multi-task learning with equal task weights. 
The dimensions of hidden layers are $\{1024\times5,40,1024\}$.
After training, $40$-dimensional BNFs for test sets are extracted and evaluated by the ABX task. DPGMM is implemented using tools developed by \cite{chang2013parallel}. DNN-BNF training is implemented using Kaldi \texttt{nnet1} recipe \cite{povey2011kaldi}.

For the systems employing FHVAE-based speaker-invariant features,  input features to DPGMM are reconstructed MFCCs $\{\bm{\hat{x}}\}$ with s-vector unification and further appended by $\Delta$+$\Delta\Delta$. 
The representative speaker is selected from the $12$ speakers mentioned in Section \ref{subsec:repre_spk}. 
The numbers of clustering iterations for the three languages are $80,80$ and $1400$. 
DNN-BNFs are trained with either reconstructed MFCCs $\{\bm{\tilde{x}}\}$ or latent segment variables $\{\bm{z_1}\}$. The extraction of $\{\bm{\tilde{x}}\}$  is slightly different from $\{\bm{\hat{x}}\}$. During the  inference of $\{\bm{\tilde{x}}\}$ for  training sets, s-vector unification is not applied;  during the inference for  test sets, s-vector unification is applied within every test subset with a subset-specific $\bm{\mu_2^*}$. 
The reason is that  DNN-BNFs  trained with $\{\bm{\tilde{x}}\}$ were found to outperform those trained with $\{\bm{\hat{x}}\}$.
The DNN-BNF mentioned here has
the same structure and loss function as that in the baseline system.


\section{Results and analyses}

\subsection{Effectiveness of reconstructed MFCCs}
\begin{figure}[t]
    \centering
    \includegraphics[width=0.95\linewidth]{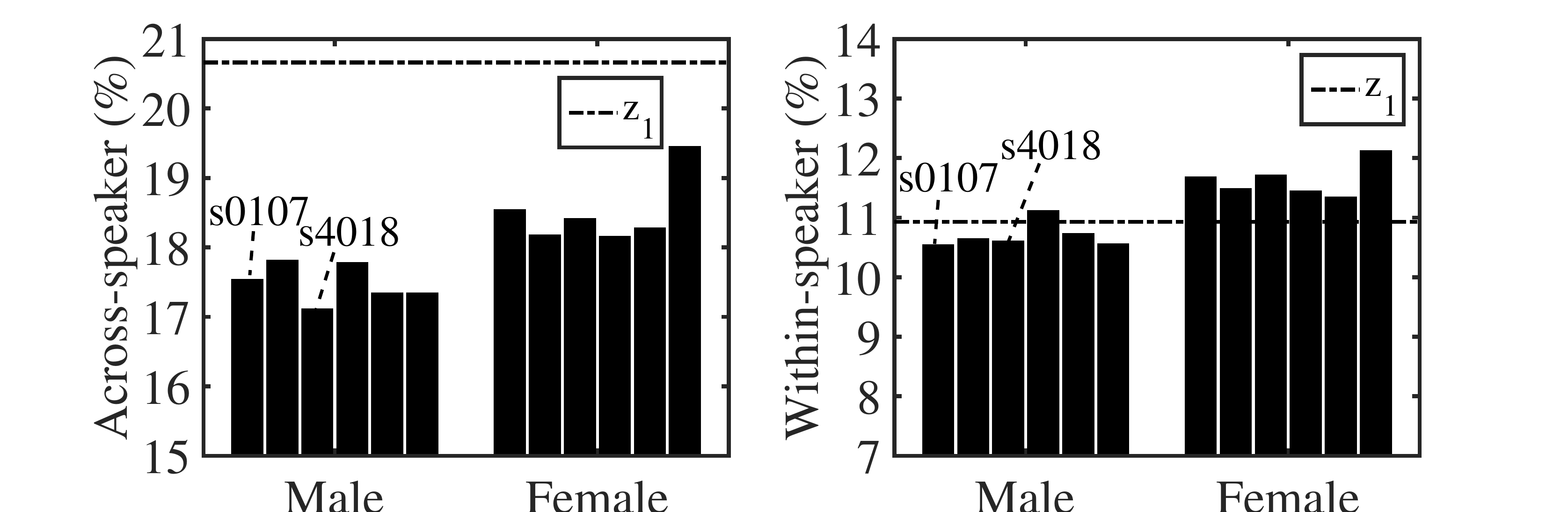}

    \caption{ABX error rates (\%) on $\bm{\hat{x}}$ using s-vector unification with different representative speakers (Avg. over languages)}
    \label{fig:recon_abx}
\end{figure}

ABX error rates on the $12$ groups of reconstructed MFCCs $\{\bm{\hat{x}}\}$ using s-vector unification is shown in Figure \ref{fig:recon_abx}. Each group 
is presented as a bar inside a bar graph.
The reference line denotes ABX error rate on latent segment variables $\{\bm{z_1}\}$.
It can be observed that,  $\{\bm{\hat{x}}\}$ outperform $\{\bm{z_1}\}$ in across-speaker condition  regardless of choosing any of the $12$ speakers as the representative. In within-speaker condition, $\{\bm{\hat{x}}\}$ perform slightly better than $\{\bm{z_1}\}$ in most of the male  cases, and are worse in all  female  cases. Further studies are needed to explain why male speakers are more suitable than females for s-vector unification.
\subsection{DNN-BNFs trained with reconstructed MFCCs}
Experimental results of the baseline DNN-BNF system and systems adopting FHVAE-based speaker-invariant features  are summarized in Table \ref{tab:results}.
The second and third columns of IDs $\X1$-$\X6$  denote inputs to DNN-BNF training and DPGMM clustering, respectively. 
`Orig.' denotes original MFCCs without reconstruction. `$\bm{\hat{x}}$-s0107/-s4018' denotes reconstructed MFCCs with representative speaker s0107 or s4018.
Here, $\bm{\hat{x}}$-s4018 is used to  represent the ideal case as s4018 performs the best among the $12$ speakers in across-speaker condition (see Figure \ref{fig:recon_abx}).
$\bm{\hat{x}}$-s0107 represents the general case as s0107 performs moderately
among  the male speakers. 
The system exploiting a  Cantonese ASR for fMLLR estimation \cite{Feng2018exploiting} is denoted as `CA-Sup'. From this Table, several observations can be made:

(1) The comparison between  baseline and \X1 \& \X2 shows that without improving frame labels, the DNN-BNF model trained with  $\{\bm{\tilde{x}}\}$ or $\{\bm{z_1}\}$ outperforms that trained with raw MFCCs, especially in across-speaker condition.

(2) The reconstructed MFCC features $\{\bm{\hat{x}}\}$ significantly outperform original MFCCs in DPGMM frame labeling. 
In the ideal case  where the representative speaker `s4018'  is  selected, by comparing \X5 and \X1,  frame labeling based on $\{\bm{\hat{x}}\}$ contributes to  $13.5\%$ and $6.9\%$ relative ABX error rate reductions in across- and within-speaker conditions, compared to that based on original MFCCs. 
In the general case where `s0107' is selected, by comparing \X3 and \X1, the relative error rate reductions are  $11.3\%$ and $4.7\%$ in across- and within-speaker conditions.
The results demonstrate the importance of  applying FHVAE-based speaker-invariant features in frame labeling.


(3) Our best system \X5 achieves $2.4\%$ and $0.6\%$ absolute ($17.3\%$ and $7.3\%$ relative) ABX error rate reductions compared to the baseline DNN-BNF system in across- and within-speaker conditions. The error rate reductions are attributed to better frame labeling and more speaker-invariant input features.
As can be seen from baseline, \X1 and \X5, the improvement in frame labeling is more prominent than that in input features.
Compared to system CA-Sup in which out-of-domain transcribed data are exploited,  \X5 is slightly better in within-speaker condition while slightly inferior in across-speaker condition.  


We also compare the effectiveness of our proposed approaches with  \cite{chen2017multilingual}, in which VTLN was adopted to improve frame labeling. As seen in Table \ref{tab:results}, in across-speaker  condition, while our baseline system is inferior to their baseline (MFCC), our best system consistently  outperforms their system MFCC+VTLN  in all test subsets. 
In within-speaker condition, our proposed approaches also achieve better performance.
The comparison shows that  FHVAE-based speaker-invariant feature learning is more effective than VTLN in improving the quality of frame labels and the robustness of subword modeling.
\section{Conclusions}

This paper presents a study on improving the quality of frame labels for unsupervised subword modeling without any out-of-domain resources.
Frame labels are generated by clustering towards speaker-invariant features learned from FHVAEs.
The speaker-invariant features are further fed as inputs to DNN-BNF training.
Experiments conducted on ZeroSpeech 2017 show that our proposed approaches achieve $2.4\%/0.6\%$ absolute  ABX error rate reductions in across-/within-speaker conditions, compared to the baseline without applying FHVAEs.
Compared with a DNN-BNF system in which out-of-domain transcribed data are used for speaker adapted feature learning, our approaches perform slightly better in within-speaker condition while slightly worse in across-speaker condition.
Our approaches significantly outperform VTLN in improving the quality of frame labels and the robustness of subword modeling.

\section{Acknowledgements}

This research is partially supported by the Major Program of National Social Science Fund of  China (Ref:13\&ZD189),
a GRF project grant (Ref: CUHK 14227216) from Hong Kong Research Grants Council  and a direct grant from CUHK Research Committee.

\bibliographystyle{IEEEtran}

\bibliography{mybib}


\end{document}